\documentclass[traditabstract]{aa} 
\usepackage{graphicx}
\usepackage{txfonts}
\usepackage{color}
\newcommand\kms{{\rm\,km\,s^{-1}}}
\newcommand\msun{{\rm\,M_\odot}}
\newcommand\rsun{{\rm\,R_\odot}}
\newcommand\masyr{{\rm\,mas\,yr^{-1}}}

\begin{document}

\title{4U\,1907+09: a HMXB running away from the Galactic plane}

\author{V.V.~Gvaramadze\inst{1,2,3}
\and S.~R\"{o}ser\inst{1} \and R.-D.~Scholz\inst{4} \and
E.Schilbach\inst{1}}

\institute{Astronomisches Rechen-Institut, Zentrum f\"{u}r
Astronomie der Universit\"{a}t Heidelberg, M\"{o}nchhofstr. 12-14,
69120 Heidelberg,
Germany\\\email{[roeser;elena]@ari.uni-heidelberg.de} \and
Sternberg Astronomical Institute, Moscow State University,
Universitetskij Pr. 13, Moscow 119992,
Russia\\\email{vgvaram@mx.iki.rssi.ru} \and Isaac Newton Institute
of Chile, Moscow Branch, Universitetskij Pr. 13, Moscow 119992,
Russia \and Astrophysikalisches Institut Potsdam, An der
Sternwarte 16, D-14482 Potsdam, Germany\\\email{rdscholz@aip.de}}

\date{Received 3 December 2010/ Accepted 4 February 2011}

\abstract{We report the discovery of a bow shock around the
high-mass X-ray binary (HMXB) 4U\,1907+09 using the {\it Spitzer
Space Telescope} 24 $\mu$m data (after Vela\,X-1 the second
example of bow shocks associated with HMXBs). The detection of the
bow shock implies that 4U\,1907+09 is moving through the space
with a high (supersonic) peculiar velocity. To confirm the runaway
nature of 4U\,1907+09, we measured its proper motion, which for an
adopted distance to the system of $4$ kpc corresponds to a
peculiar transverse velocity of $\simeq 160 \pm 115 \, \kms$,
meaning that 4U\,1907+09 is indeed a runaway system and supporting
the general belief that most of HMXBs possess high space
velocities. The direction of motion of 4U\,1907+09 inferred from
the proper motion measurement is consistent with the orientation
of the symmetry axis of the bow shock, and shows that the HMXB is
running away from the Galactic plane. We also present the {\it
Spitzer} images of the bow shock around Vela\,X-1 (a system
similar to 4U\,1907+09) and compare it with the bow shock
generated by 4U\,1907+09.}

\keywords{proper motions -- stars: early-type -- stars: kinematics
and dynamics -- stars: individual: 4U\,1907+09, Vela\,X-1 --
X-rays: binaries}

\maketitle

\section{Introduction}
%
\object{4U\,1907+09} is a high-mass X-ray binary (HMXB),
consisting of an X-ray pulsar (neutron star) accreting from a late
O-type supergiant companion star (Giacconi et al. \cite{gi71};
Schwartz et al. \cite{sc80}; Marshall \& Ricketts \cite{ma80};
Makishima et al. \cite{ma84}; van Kerkwijk, van Oijen \& van den
Heuvel \cite{va89}; Cox, Kaper \& Mokiem \cite{cox05}; Nespoli,
Fabregat \& Mennickent \cite{ne08}). Like other HMXBs, 4U\,1907+09
is the descendant of a massive binary, whose primary (initially
more massive) star experienced supernova explosion (van den Heuvel
\& Heise \cite{va72}). The fact that the binary has remained bound
after the supernova explosion implies either that the system lost
less than a half of its initial (pre-supernova) mass (in the case
of a symmetric supernova explosion; Boersma \cite{bo61}) or that
the kick attained by the neutron star at birth had a favorauble
orientation and magnitude (if the explosion was asymmetric; Hills
\cite{hi83}; Tauris \& Takens \cite{ta98}). In both cases, the
HMXB attains a recoil velocity, which could be large enough (i.e.
$\geq 30 \, \kms$) to classify the system as a runaway (Blaauw
\cite{bl61}). The distribution of HMXBs in the Galaxy and their
kinematic properties are consistent with the possibility that most
of them are runaways (van Oijen \cite{van89}; Chevalier \&
Ilovaisky \cite{ch98}). A similar conclusion was derived by Coe
(\cite{co05}) for HMXBs in the Small Magellanic Cloud.

The high peculiar velocities of HMXBs could be revealed directly,
via measurement of their radial velocities (van Oijen 1989) and/or
proper motions (Chevalier \& Ilovaisky 1998), or indirectly,
through the detection of the secondary attributes of runaway stars
-- the bow shocks (e.g. van Buren \& McCray 1988; van Buren,
Noriega-Crespo \& Dgani 1995; Gvaramadze \& Bomans 2008;
Gvaramadze, Kroupa \& Pflamm-Altenburg 2010a; Gvaramadze,
Pflamm-Altenburg \& Kroupa 2011). The first-ever detection of a
bow shock associated with a HMXB was reported by Kaper et al.
(1997), who discovered a very symmetric H$\alpha$ arc around the
HMXB \object{Vela\,X-1} (a system similar to 4U\,1907+09). This
discovery unambiguously showed that at least in the case of
Vela\,X-1 the system got its high space velocity owing to the
binary-supernova explosion [an alternative mechanism for the
origin of runaway stars (including binaries) is based on dynamical
three- and four-body encounters in dense star clusters; Poveda,
Ruiz \& Allen \cite{po67}; Gies \& Bolton \cite{gi86})].

Huthoff \& Kaper (\cite{hu02}) used the high-resolution {\it IRAS}
maps to search for bow shocks around 11 high-velocity HMXBs and
detected only an infrared counterpart to the already known bow
shock generated by Vela\,X-1. The paucity of bow shock-producing
HMXBs is consistent with the observational fact that only a
minority of runaway OB stars are associated with (detectable) bow
shocks (van Buren et al. \cite{va95}). The most reliable
explanation of this fact is that the majority of runaway stars are
moving through a low density, hot medium, so that the emission
measure of their bow shocks is below the detection limit or the
bow shocks cannot be formed at all because the sound speed in the
local interstellar medium (e.g. the hot gas within large-scale
bubbles around OB associations) is higher than the stellar space
velocity (Kaper, Comer\'{o}n \& Barziv \cite{ka99}; Huthoff \&
Kaper \cite{hu02}). In case of bow shocks with small emission
measure, one might expect to detect them using modern infrared
telescopes.

Motivated by the above arguments, we have undertaken a search for
bow shocks around the HMXBs from the sample of Huthoff \& Kaper
(\cite{hu02}; see their Table\,1) using the archival {\it Spitzer
Space Telescope} data. We have discovered a bow shock around the
HMXB 4U\,1907+09 and detected the already known bow shock
associated with Vela\,X-1 (Sect.\,\ref{sec:bow}). The orientation
of the bow shock generated by 4U\,1907+09 is consistent with
ejection of this HMXB from the Galactic plane. To confirm the
runaway nature of the system, we have measured its proper motion
using archival data over a total time baseline of 50 yr
(Sect.\,\ref{sec:prop}). Sect.\,\ref{sec:dis} discusses the
results obtained in the previous sections. We summarize and
conclude in Sect.\,\ref{sec:sum}.

\section{4U\,1907+09: general data}
\label{sec:data}

The HMXB 4U\,1907+09 was discovered with {\it UHURU} as a discrete
X-ray source (Giacconi et al. \cite{gi71}). The X-ray source was
subsequently identified with a highly reddened early-type star
with broad, strong H$\alpha$ emission (Schwartz et al.
\cite{sc80}). Schwartz et al. (\cite{sc80}) hypothesized that
4U\,1907+09 is a binary system composed of an OB supergiant star
and a compact object, and that accretion of the stellar wind onto
the compact object is responsible for the X-ray emission. This
hypothesis was proved by Marshall \& Ricketts (\cite{ma80}), who
revealed a binary period of $\simeq 8.38$ day in the X-ray light
curve of 4U\,1907+09, confirming that the system is a new member
of a class of HMXBs (see, e.g., Kaper \& van der Meer \cite{ka07}
for a review on these objects). Makishima et al. (\cite{ma84})
discovered a 437.5 s X-ray pulsations from 4U\,1907+09,
establishing this system as a binary X-ray pulsar.

Follow-up optical spectroscopy of 4U\,1907+09 by Cox et al.
(\cite{cox05}) showed that the companion of the X-ray pulsar is a
late O-type supergiant star (O8\,Ia-O9\,Ia; cf. van Kerkwijk et
al. \cite{va89}; Iye \cite{iy86}). This classification was refined
by Nespoli et al. (\cite{ne08}), who got the infrared spectrum of
4U\,1907+09 and derived the spectral type of O9.5\,Iab.

The distance to 4U\,1907+09 still remains uncertain (cf. van
Kerkwijk et al. \cite{va89}; Cox et al. \cite{cox05}; Nespoli et
al. \cite{ne08}).

Using the optical photometry from Schwarts et al. (\cite{sc80}),
$V=16.37\pm0.02$ mag, $B-V=3.17\pm 0.06$ mag, and adopting the
intrinsic $(B-V)_0$ colour of an O9.5\,I star of $-0.26$ mag
(Martins \& Plez \cite{ma06}), one finds the interstellar
reddening towards 4U\,1907+09 $E(B-V)=3.43\pm 0.06$ mag. Then
taking $A_V =3.1E(B-V)$, one obtains $A_V \simeq 10.6\pm 0.2$,
which for $M_V =-6.34\pm 0.45$ mag (Martins \& Plez \cite{ma06})
leads to a distance $d\simeq 2.6^{+0.7} _{-0.5}$ kpc.

Similarly, using $J$ and $K_s$ magnitudes of 4U\,1907+09
($10.00\pm 0.02$ and $8.77 \pm 0.02$, respectively) from 2MASS
(Skrutskie et al. \cite{sk06}), $(J-K_s)_0 =-0.21$ mag and
$M_{K_{\rm s}} =-5.52\pm 0.20$ mag from Martins \& Plez
(\cite{ma06}), and adopting the extinction law from Rieke \&
Rebofsky (\cite{ri85}), one finds $A_{K_{\rm s}} = 0.66[(J-K_{\rm
s}) -(J-K_{\rm s})_0] = 0.95\pm 0.02$\footnote{We also tried to
estimate $A_{K_{\rm s}}$ using the {\it Spitzer} photometry of
4U\,1907+09 and the Stellar Performance Estimation Tool
(http://ssc.spitzer.caltech.edu/warmmission/propkit/pet/starpet/
index.html), but found that the fluxes predicted by this tool
(based on Kurucz-Lejeune atmospheric models) cannot be reconcilled
with the observed ones.}, suggesting a distance of $\simeq
4.7^{+2.7} _{-1.7}$ kpc. The large discrepancy between the
distance estimates based on the visual and near infrared
photometry could be understood if the reddening towards
4U\,1907+09 is anomalous, i.e. the total-to-selective absorption
ratio $R_V < 3.1$ (cf. Cox et al. \cite{cox05}).

Using a similar procedure and adopting the intrinsic colours and
absolute magnitudes from Wegner (\cite{we94}, \cite{we06}),
Nespoli et al. (\cite{ne08}) derived a distance to 4U\,1907+09 of
$2.8^{+5.0} _{-1.8}$ kpc, where the quoted errors stem mainly from
the uncertainty in the absolute magnitudes.

A distance $d=2.1-2.6$ kpc was derived by van Kerkwijk et al.
(\cite{va89}) and Cox et al. (\cite{cox05}) from interstellar
absorption of sodium in the spectrum of 4U\,1907+09, while the
detection of a high velocity ($\sim 60 \, \kms$) component in the
K\,{\sc i} absorption line in the line-of-sight towards
4U\,1907+09 suggests a distance of $\geq 4$ kpc (Cox et al.
\cite{cox05}). The latter estimate is based on the model of
Galactic rotation by Brand \& Blitz (\cite{br93}), which assumes
that, to first order, the Galactic velocity field is axisymmetric.
Note, however, that 4U\,1907+09 is located in the direction of the
Sagittarius arm tangent region and that our line-of-sight towards
this HMXB passes through the spiral arm over a distance of several
kpc. One cannot, therefore, exclude that random (non-circular)
motions (up to several tens of $\kms$) caused by outflows from
star-forming regions could contribute to the origin of the
high-velocity components in the K\,{\sc i} absorption line,
thereby affecting the kinematic distance estimate.

Finally, we note that the interstellar extinction cannot be
constrained by using the empirical relationships between $A_V$ and
the interstellar column density $N_{\rm H}$ (derived from
modelling the X-ray spectrum of the neutron star), because the
total column density towards 4U\,1907+09 contains a variable
contribution from the circumstellar material of a magnitude
comparable to that of the interstellar one (Cook \& Page
\cite{co87}; in 't Zand, Strohmayer \& Baykal \cite{in97}; Rivers
et al. \cite{ri10}).

As a compromise between the different distance estimates, we adopt
a distance to 4U\,1907+09 of 4 kpc in the following.

\section{4U\,1907+09: bow shock} \label{sec:bow}

\begin{figure*}
\includegraphics[width=18cm]{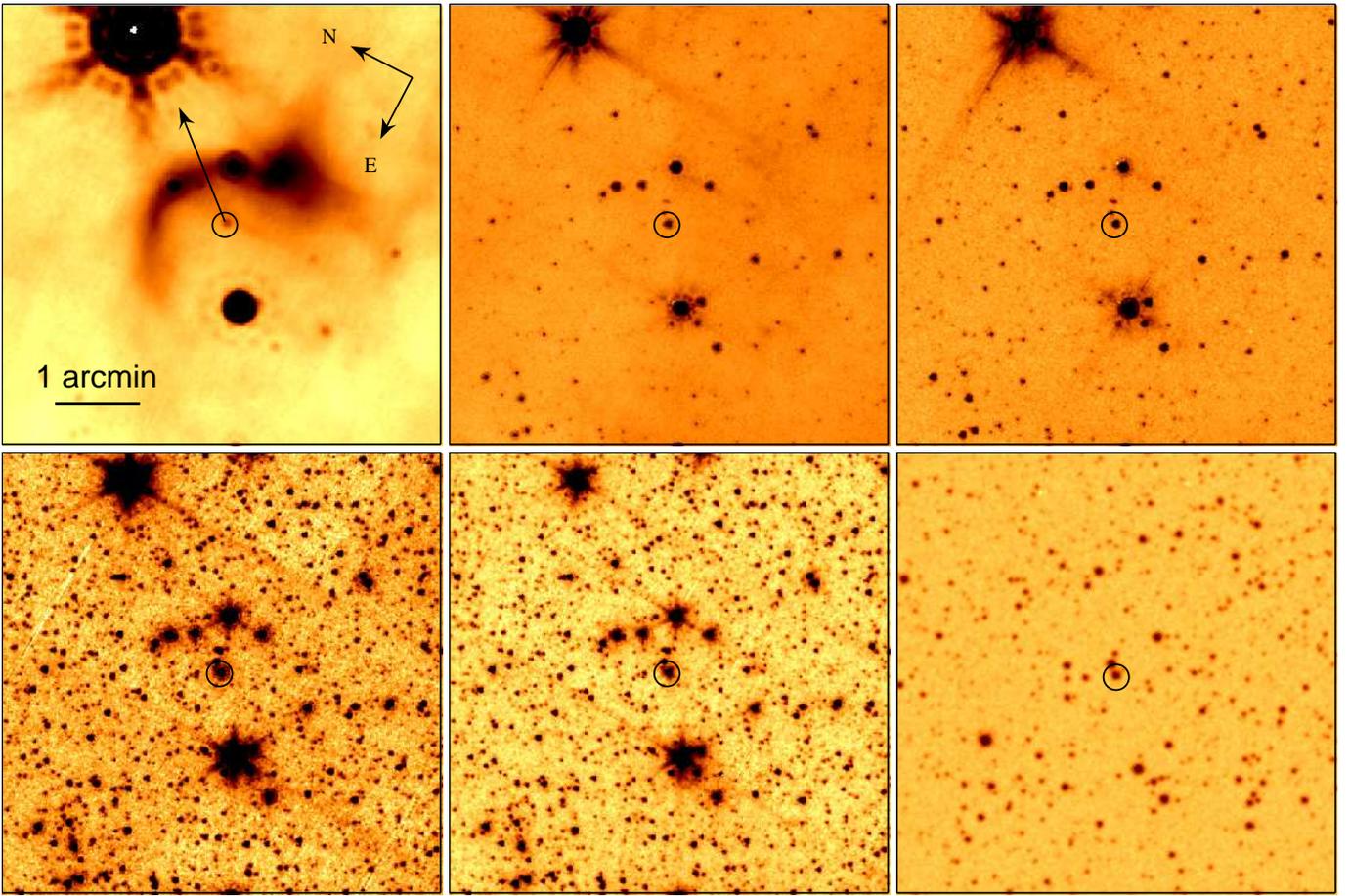}
\centering \caption{From left to right, and from top to bottom:
{\it Spitzer} MIPS $24\,\mu$m, IRAC 8, 5.8, 4.5 and 3.6 $\mu$m,
and DSS-II (red band) images of the field around the HMXB
4U\,1907+09. The arrow shows the direction of peculiar velocity of
4U\,1907+09, as suggested by the proper motion measurement (see
text for details). A bright infrared source to the north is the
maser source IRAC\,19071+0946. In all panels the position of
4U\,1907+09 is indicated by a circle.} \label{fig:bow}
\end{figure*}
\begin{figure*}
\sidecaption
\includegraphics[width=14cm]{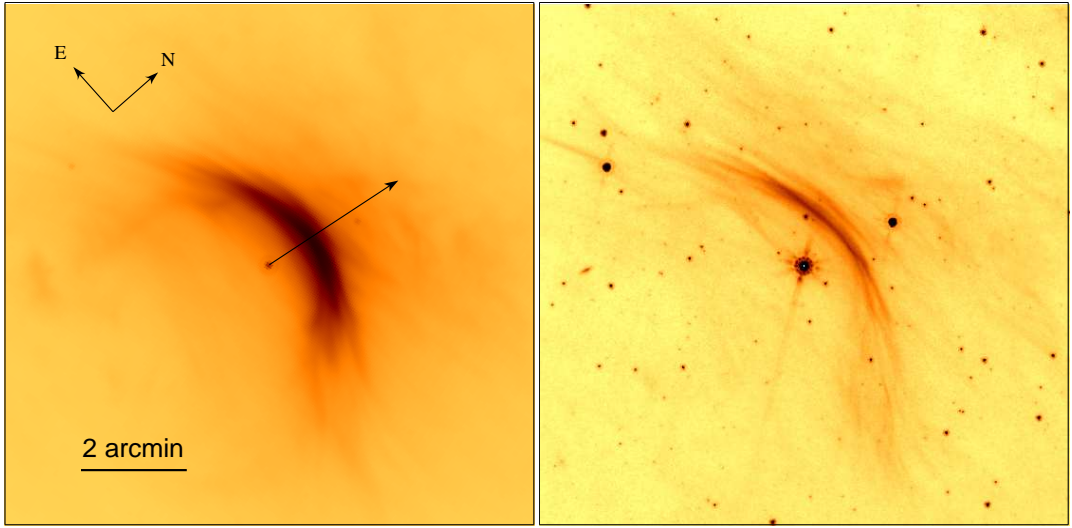}
\centering \caption{{\it Spitzer} MIPS $24\,\mu$m (left) and IRAC
$8\,\mu$m images of the bow shock around the HMXB Vela\,X-1. The
arrow shows the direction of peculiar velocity of Vela\,X-1 (see
Sect.\,\ref{sec:shock} for details).} \label{fig:Vela}
\end{figure*}

To search for bow shocks around HMXBs from the sample of Huthoff
\& Kaper (\cite{hu02}), we utilized the imaging data from the {\it
Spitzer Space Telescope} archive, obtained with the Multiband
Imaging Photometer for {\it Spitzer} (MIPS; Rieke et al.
\cite{ri04}) and the Infrared Array Camera (IRAC; Fazio et al.
\cite{fa04}). The publicly available data cover the fields
containing 5 of the 11 HMXSs (namely, \object{GX\,301-2},
4U\,1907+09, \object{Cyg\,X-1}, Vela\,X-1 and \object{V615\,Cas}).
Visual inspection of the MIPS and IRAC images results in a
discovery of a bow shock generated by 4U\,1907+09.

Figure\,\ref{fig:bow} shows the MIPS $24\,\mu$m, IRAC 8, 5.8, 4.5
and 3.6 $\mu$m, and DSS-II (red band) images\footnote{The MIPS and
IRAC images were obtained within the framework of the 24 and 70
Micron Survey of the Inner Galactic Disk with MIPS (MIPSGAL; Carey
et al. \cite{ca09}) and the Galactic Legacy Infrared Mid-Plane
Survey Extraordinaire (GLIMPSE; Benjamin et al. \cite{be03}),
respectively.} of the field around the HMXB 4U\,1907+09. Like many
other bow shocks discovered with {\it Spitzer} (e.g. Gvaramadze et
al. \cite{gv10a},b, \cite{gva11}; Gvaramadze \& Gualandris
\cite{gv11}), the bow shock associated with 4U\,1907+09 is only
visible at $24 \, \mu$m (although the IRAC\, 5.8 and 8 $\mu$m
images show the gleam of emission possibly associated with the bow
shock). The bow shock has a clear arcuate shape with the apex at
$\simeq 0\farcm7$ from 4U\,1907+09, which at the distance to the
system of 4 kpc corresponds to the linear separation $R_{\rm obs}
\simeq 0.8$ pc (see Sect.\,\ref{sec:shock}).

We also detected the already known bow shock associated with
Vela\,X-1 [see Kaper et al. (\cite{ka97}) and Huthoff \& Kaper
(\cite{hu02}) for the H$\alpha$ and {\it IRAS} 60 $\mu$m images of
the bow shock, respectively]. Unlike 4U\,1907+09, the bow shock
produced by Vela\,X-1 is visible not only at 24 $\mu$m, but also
at all four IRAC bands, i.e. at 3.6, 4.5, 5.8 and $8 \, \mu$m. In
Fig.\,\ref{fig:Vela} we present for the first time the MIPS
24\,$\mu$m and IRAC 8\,$\mu$m images of the bow shock (Program
Id.: 30088, PI: A.Noriega-Crespo)\footnote{Note that the bow shock
produced by Vela\,X-1 was independently observed with {\it
Spitzer} by R.Iping (Program Id.: 30174).} showing its fine
structure (most prominent at 8\,$\mu$m) similar to that detected
in H$\alpha$ by Kaper et al. (\cite{ka97}). Using the IRAC
$8\,\mu$m image of the bow shock, we estimated $R_{\rm obs} \simeq
0.5 \, d_{1.9}$ pc, where $d_{1.9}$ is the distance to Vela\,X-1
in units of 1.9 kpc (Sadakane et al. \cite{sa85}).

The symmetry axis of a bow shock generated by a supersonically
moving star reflects the direction of motion of the star with
respect to the ambient medium, which also could be in motion
relative to the local standard of rest at the location of the
star. For example, high-velocity outflows driven from young star
clusters by the collective effect of stellar winds can create bow
shocks around (low-velocity) stars in the cluster's halo (e.g.,
bow shocks around LL Ori and several other stars in the Orion
Nebula, all of which are faced towards the central Trapezium
cluster; Bally, O'Dell \& McCaughrean \cite{ba00}). For runaway
stars located far from star-forming regions and their associated
gas outflows (caused by stellar winds and supernovae), one can
assume that the peculiar velocity of the ambient medium is
negligible compared with the space velocity of the star, meaning
that the symmetry axis of the bow shock coincides well with the
direction of stellar motion. In this case, the orientation of the
bow shock can be used to back-trace the trajectory of the star to
the parent cluster even for those stars whose proper motions
cannot be measured with a high confidence (Gvaramadze \& Bomans
\cite{gv08}; Gvaramadze et al. \cite{gv10a},b, \cite{gva11}).

\begin{figure}
\includegraphics[width=9cm]{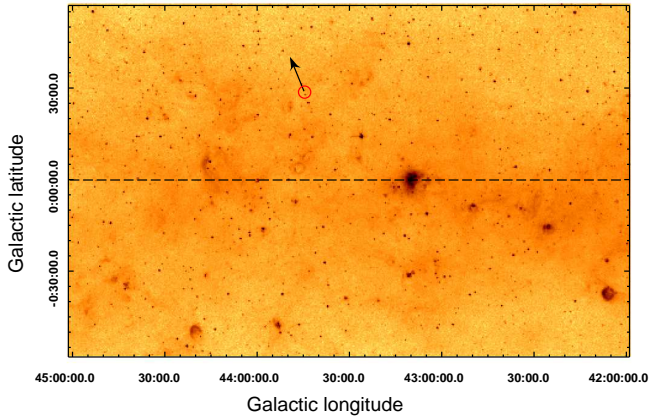}
\centering \caption{$3\degr \times 2\degr$ {\it MSX} $8.3 \,\mu$m
image of the Galactic plane (shown by a dashed line) centred at
$l=43\fdg5, b=0\degr$, with the position of the HMXB 4U\,1907+09
indicated by a circle. The arrow shows the direction of motion of
4U\,1907+09. The orientation of this figure is the same as for
Fig.\,\ref{fig:bow}. The brightest nebula on the image is the
background star-forming region W\,49A.} \label{fig:run}
\end{figure}

Figure\,\ref{fig:run} shows the {\it Midcourse Space Experiment}
({\it MSX}) satellite (Price et al. \cite{pr01}) image of a
$3\degr \times 2\degr$ region of the Galactic plane (centred at
$l=43\fdg5, b=0\degr$), with the position of 4U\,1907+09 indicated
by a circle ($l=43\fdg7436 ,b=0\fdg4753$). The symmetry axis of
the bow shock around 4U\,1907+09 (see Fig.\,\ref{fig:bow}, whose
orientation is the same as for Fig.\,\ref{fig:run}) suggests that
this HMXB is moving away from the Galactic plane. To prove
unambiguously that the origin of the bow shock is due to the high
space velocity of 4U\,1907+09, we searched for existing proper
motion measurements for this system using the Vizier catalogue
access tool\footnote{http://webviz.u-strasbg.fr/viz-bin/VizieR}.
We found three catalogues giving proper motions for 4U\,1907+09,
USNO-B1 (Monet et al. \cite{mo03}), UCAC3 (Zacharias et al.
\cite{za10}) and PPMXL (R\"oser, Demleitner \& Schilbach
\cite{ro10}). All three give huge proper motions, which, if real,
would correspond to the peculiar velocity of several thousands of
$\kms$. We discovered, however, that USNO-B1 has two entries
separated by less than $2\arcsec$ at the position of 4U\,1907+09
which (possibly) originate from mismatching of plates from
different epochs and causes the huge proper motion. PPMXL followed
the cross-matching of USNO-B1, hence the star is one of the 10\%
(R\"oser et al. \cite{ro10}) double entries in PPMXL. Also, it
turns out that the UCAC3 proper motion is marked ``with doubts"
because it relies on ``less than two good matches". Thus, an
individual determination of the proper motion of 4U\,1907+09
became needful.

\section{4U\,1907+09: proper motion measurement}
\label{sec:prop}

With its intermediate optical magnitudes, 4U 1907$+$09 is not
saturated on photographic Schmidt plates (see
Fig.\,\ref{fig:bow}), providing different epoch observations over
five decades, and it is well measured in the CCD-based Carlsberg
Meridian Catalog 14
(CMC14)\footnote{http://www.ast.cam.ac.uk/cmt/cmc14.html}.
However, there are several difficulties in the field around the
target: 1) overlap of star images, changing already with the
different optical passbands and leading to a completely different
appearance of the field in the near infrared due to strong
reddening, 2) absence of reference galaxies for direct absolute
proper motion measurement, 3) different location of the target on
the Schmidt plates resulting in different residual geometric
distortions after any full Schmidt plate calibration.

We collected all available Schmidt plate data, POSS-1 blue (O) and
red ($E$), POSS Quick $V$, and POSS-2 $B, R$, and $I$, altogether
9 plates, and extracted small 15$\times$15 arcmin$^2$ FITS images
from the
DSS\footnote{http://archive.stsci.edu/cgi-bin/dss\_plate\_finder}.
In case of the POSS-1 red (E) plate we also extracted the FITS
file from the SuperCOSMOS Sky
Survey\footnote{http://www-wfau.roe.ac.uk/sss/pixel.html} (Hambly
et al. \cite{ha01}), which we preferred because of its higher
resolution. To overcome the above mentioned difficulties, we used
14 reference stars (of magnitudes similar to those of 4U\,
1907+09) closely distributed around the target within about 3
arcmin for a new local astrometric calibration of the Schmidt
plates with respect to the CMC14 positions of the reference stars.
The astrometric measurements and transformations from the Schmidt
plates to the CMC14 system were carried out using the ESO SkyCat
Tool\footnote{http://archive.eso.org/cms/tools-documentation/skycat}.
To avoid colour-dependent effects, we finally preferred to combine
only the $E, R$ and $V$ plates with the CMC14 $(r')$-band data.

The resulting proper motion from simple linear fitting of the
target positions at five epochs (1952.395, 1983.611, 1987.411,
1992.586, 2001.956) is $\mu _{\alpha} \cos \delta = -4.1 \pm 5.5
\, \masyr, \mu _{\delta} = 4.9 \pm 4.4 \, \masyr$. For the
absolutisation of this relative proper motion we investigated the
proper motions of the reference stars as given in the PPMXL
catalogue (R\"{o}ser et al. \cite{ro10}), which is on the
International Celestial Reference System. The individual PPMXL
proper motions of the 14 stars are all smaller than $10 \,
\masyr$, and their mean PPMXL proper motion is $\mu _{\alpha} \cos
\delta = -3.6 \pm 3.0 \, \masyr$, $\mu _{\delta} = -3.4 \pm 3.0 \,
\masyr$. The quoted errors are conservative estimates including
the systematic PPMXL uncertainty of 1 to $2 \, \masyr$ in any
small field ($\sim$1 square degree) of the sky due to the small
number density of {\it Hipparcos} stars (R\"{o}ser et al.
\cite{ro10}). Taking into account the mean proper motion of the 14
reference stars we then computed an absolute proper motion of
4U\,1907+09: $\mu _{\alpha} \cos \delta = -7.7 \pm 6.3 \, \masyr,
\mu _{\delta} = 1.5 \pm 5.3 \, \masyr$, or in Galactic
coordinates:
\begin{eqnarray}
\mu_l = -2.2 \pm 5.5 \, \masyr \, , \, \mu_b = 7.5 \pm 6.1 \,
 \masyr \, . \nonumber
\label{eqn:vel}
\end{eqnarray}

\section{Discussion}
\label{sec:dis}

\subsection{4U\,1907+09 as a runaway HMXB}
\label{sec:run}

To convert the observed proper motion of 4U\,1907+09 into the
transverse peculiar velocity, we used the Galactic constants $R_0
= 8.0$ kpc and $\Theta _0 =240 \, {\rm km} \, {\rm s}^{-1}$ (Reid
et al. 2009) and the solar peculiar motion $(U_{\odot} , V_{\odot}
, W_{\odot})=(10.0, 11.0, 7.2) \, {\rm km} \, {\rm s}^{-1}$
(McMillan \& Binney \cite{mc10})]. We found the components of the
peculiar velocity:
\begin{equation}
v_l = 59\pm 104 \, \kms , v_b =150\pm 116 \, \kms \nonumber
\label{eqn:vel}
\end{equation}
(here we assumed that the errors are due to errors in the proper
motion measurement). Taken at face value, these velocity
components suggest that 4U\,1907+09 is running away from the
Galactic plane in the direction consistent with the orientation of
the symmetry axis of the bow shock (see Figs\,\ref{fig:run} and
\ref{fig:bow}; recall that the orientation of both figures is the
same).

By tracing back the trajectory of 4U\,1907+09 and assuming that
this HMXB originated in the Galactic plane, we inferred that the
system was ejected from the region around $l=43\fdg5, b=0\degr$.
We did not find, however, in this region (of radius of $1\degr$)
any known young ($\la 10$ Myr) star cluster or association located
at a distance comparable to that of 4U\,1907+09. (The only young
star-forming region in this area is the background one
\object{W\,49A}.) This could simply be owing to the huge
interstellar extinction in the Galactic plane. Another possibility
is that the progenitor massive binary was dynamically expelled
from the parent cluster well before the primary exploded as a
supernova and that the system changed the direction of its motion
because of a kick caused by the supernova explosion (cf.
Pflamm-Altenburg \& Kroupa \cite{pf10}). In this case, the parent
cluster could be located at several degrees from the region
inferred from the proper motion measurement and orientation of the
bow shock.

We also searched for the parent diffuse supernova remnant in the
same area of the Galactic plane using the catalogue of Galactic
supernova remnants by Green
(\cite{gr09})\footnote{http://www.mrao.cam.ac.uk/surveys/snrs/}.
The only supernova remnant listed in the catalogue in this area is
the background one \object{W\,49B}. The non-detection, however, is
not surprising because the kinematic age of 4U\,1907+09 (i.e. the
time elapsed since the supernova explosion), $t_{\rm kin} = d\sin
b /v_b \sim 10^5$ yr, is comparable to the typical lifetime of
diffuse supernova remnants (e.g. Shull, Fesen \& Saken
\cite{sh89}).

\subsection{Origin of a high peculiar velocity of 4U\,1907+09}
\label{sec:vel}

\subsubsection{Symmetric supernova explosion}
\label{sec:velsym}

Boersma (\cite{bo61}) showed that a post-supernova binary remains
bound if the mass of the supernova ejecta, $M_{\rm ej}$, comprises
less than a half of the system's initial mass, i.e. $M_{\rm ej} <
(1/2)(M_1 +M_2)$, where $M_1$ and $M_2$ are the pre-supernova
masses of the primary and secondary stars, respectively. In
response to the instant mass loss caused by the supernova
explosion, the binary system recoils with a velocity, given by
(Gott, Gunn \& Ostriker \cite{go70}; Iben \& Tutukov \cite{ib97})
\begin{equation}
v_{\rm sym} =\left({GM_2 \over a_{\rm pre-SN}}\right)^{1/2}
\left({M_2 \over M_1 +M_2}\right)^{1/2} {M_1 -M_{\rm co} \over M_2
+M_{\rm co}} \, , \label{eqn:velsym}
\end{equation}
where $G$ is the gravitational constant, $a_{\rm pre-SN}$ is the
pre-supernova semimajor axis of the binary, and $M_{\rm co} =M_1
-M_{\rm ej}$ is the mass of the compact object (a neutron star or
black hole). In case of 4U\,1907+09, the stellar supernova remnant
is a neutron star and we assume that it has a canonical mass
$M_{\rm co} =1.4 \msun$.

It follows from Eq.\,(\ref{eqn:velsym}) that the HMXB achieves the
highest peculiar velocity if the pre-supernova binary was as tight
as possible, i.e. if the secondary star of radius $R_2$ was close
to filling its Roche lobe, $R_2 \sim R_{\rm L}$, where $R_{\rm L}$
is the radius of the Roche lobe, given by (Eggleton \cite{eg83})
\begin{equation}
R_{\rm L} = {0.49a_{\rm pre-SN}q^{2/3} \over 0.6q^{2/3} +\ln
(1+q^{1/3})} \, ,
\end{equation}
and $q=M_2 /M_1$. Eq.\,(\ref{eqn:velsym}) also shows that the
larger $M_{\rm ej} =M_1 -M_{\rm co}$ the higher $v_{\rm sym}$.
Assuming that at the moment of supernova explosion the secondary
was a $26 \, \msun$ main-sequence star (cf. Cox et al.
\cite{cox05}) of radius
\begin{equation}
R_2 = 0.8(M_2 /\msun)^{0.7}\rsun \label{eqn:rad}
\end{equation}
(Habets \& Heintze \cite{ha81}) and adopting the maximum
pre-supernova mass of the exploding star of $17 \, \msun$ [see
Fig.\,6 of Meynet \& Maeder (\cite{me03})], one finds from
Eqs\,(\ref{eqn:velsym})-(\ref{eqn:rad}) that $a_{\rm pre-SN}
\simeq 2.4R_{\rm L} \simeq 20 \, \rsun$ and the maximum possible
systemic velocity $v_{\rm sym} ^{\rm max} \simeq 230 \, \kms$. The
latter figure, when compared with the total peculiar (transverse)
velocity of 4U\,1907+09, $v_{\rm tr} =(v_l ^2 +v_b ^2)^{1/2}
\simeq 160\pm 115 \, \kms$ [see Eq.\,(\ref{eqn:vel})], means that
the symmetric supernova explosion could be responsible for the
origin of the high space velocity of 4U\,1907+09.

The maximum recoil velocity would be smaller if at the moment of
supernova explosion the secondary star already entered the
supergiant stage of evolution. Adopting $R_{\rm L} \sim R_2 =26 \,
\rsun$ (Cox et al. \cite{cox05}), one finds $v_{\rm sym} ^{\rm
max} \simeq 130 \, \kms$. Still, the ``measured" peculiar velocity
of 4U\,1907+09 within the 1$\sigma$ error bar is consistent with
the maximum possible recoil velocity. Note, however, that $a_{\rm
pre-SN}$ is related to the post-supernova semimajor axis, $a_{\rm
post-SN}$, through the relationship:
\begin{eqnarray}
a_{\rm pre-SN} = \left({M_2 +2M_{\rm co} -M_1 \over M_{\rm co}
+M_2}\right) a_{\rm post-SN} \, , \nonumber \label{eqn:semi}
\end{eqnarray}
from which it follows that $a_{\rm pre-SN} \simeq 24 \, \rsun$ if
one uses the stellar masses adopted above and the present
semimajor axis of 4U\,1907+09 of $\simeq 50 \, \rsun$. This
estimate suggests that at the moment of supernova explosion the
secondary star was still on the main-sequence and left it later
on.

\subsubsection{Asymmetric supernova explosion}
\label{sec:velasym}

The recoil velocity could be higher if the supernova explosion was
asymmetric and the stellar remnant received a kick at birth. In
this case, the systemic velocity of the post-supernova binary
depends also on the magnitude and orientation of the kick, and is
given by (Stone \cite{st82}):
\begin{equation}
v_{\rm asym} = \left(v_{\rm sym} ^2 -2v_{\rm sym} v_{\rm k} \cos
\psi + v_{\rm k} ^2 \right)^{1/2} \, , \label{eqn:velasym}
\end{equation}
where $v_{\rm sym}$ is given by Eq.\,(\ref{eqn:velsym}), $v_{\rm
k} = M_{\rm co} w/(M_{\rm co} +M_2 )$, $w$ is the kick velocity,
and $\psi$ is the angle between the kick velocity and the
pre-explosion relative orbital velocity [note that
Eq.\,(\ref{eqn:velasym}) neglects the mass loss from the secondary
star owing to stripping and ablation of the stellar material
caused by the passing supernova shock wave; see Tauris et al.
(\cite{ta99}) for details].

It follows from Eq.\,(\ref{eqn:velasym}) that $v_{\rm asym}$ is
maximum if the kick velocity attained by the stellar supernova
remnant was oriented in the opposite direction to the relative
orbital velocity, i.e. $\cos \psi =-1$. In this case
\begin{equation}
v_{\rm asym} ^{\rm max} = v_{\rm sym} ^{\rm max} + M_{\rm co}
w/(M_{\rm co} +M_2).
\label{eqn:full}
\end{equation}
Eq.\,(\ref{eqn:full}) shows that for HMXBs comprising a neutron
star, i.e. for systems with $M_{\rm co} /(M_{\rm co} +M_2) << 1$,
the effect of asymmetry of supernova explosion on the systemic
velocity is modest and could be substantial (larger than several
tens of $\kms$) only for extremely large kick velocities, $w\geq
1000 \, \kms$. It is generally believed that neutron stars can
obtain kick velocities of this magnitude. This belief got a
support after Chatterjee et al. (\cite{ch05}) derived a space
velocity of $\sim 1000 \, \kms$ for the pulsar \object{PSR
B1508+55} by measuring its proper motion and parallax. Still, the
existence of a population of extremely high-velocity neutron stars
can be explained without exploiting the idea of asymmetric
supernova explosion -- they simply could be the remnants of
massive hypervelocity runaway stars (Gvaramadze \cite{gv07},
\cite{gv09}; Gvaramadze, Gualandris \& Portegies Zwart
\cite{gva08}; \cite{gva09}).

Assuming that the neutron star in 4U\,1907+09 attained a kick
velocity of $200-400 \, \kms$ at birth (typical of radio pulsars;
e.g. Hobbs et al. \cite{ho05}) and adopting $M_{\rm co} = 1.4 \,
\msun$ and $M_2 = 26 \, \msun$ (Cox et al. \cite{cox05}), one
finds from Eq.\,({\ref{eqn:full}) that the asymmetry of the
supernova explosion can increase the systemic velocity only by
$\simeq 10-20 \, \kms$ (provided that the kick velocity was of
favourable orientation; see above). We therefore conclude that the
asymmetric supernova explosion did not contribute significantly to
the space velocity of 4U\,1907+09, unless the kick velocity was
larger than $1000 \, \kms$.

\subsection{4U\,1907+09 and Vela\,X-1}
\label{sec:shock}

\begin{table*}
  \caption{Summary of parameters related to 4U\,1907+09 and Vela\,X-1.
    }
  \label{tab:wind}
  \renewcommand{\footnoterule}{}
  \begin{tabular}{ccccccccccc}
      \hline
      \hline
      HMXB & Spectral type & $d$ (kpc) & $v_\ast \, (\kms )$ & $\dot{M} \, (10^{-6} \, \msun \, {\rm yr}^{-1})$ & $v_{\rm w} \, (\kms )$ & $R_{\rm obs}$ (pc) & $n \, ({\rm cm}^{-3} ) $ \\
       \hline
       4U\,1907+09 & O9.5\,Iab$^{a}$ & 4 & $160$ & 6$^{b}$ & 1690$^{b}$ &  0.8 & 2 \\
       \hline
      Vela\,X-1 & B0.5\,Iab$^{c}$ & 1.9 & 50 & 2$^{d}$ & 1100$^{e}$ & 0.5 & 10 \\
      \hline
       \end{tabular}
       \tablefoot{
 \tablefoottext{a}{Nespoli et al. \cite{ne08}.}
 \tablefoottext{b}{Cox et al. \cite{cox05}.}
 \tablefoottext{c}{Conti \cite{co78}.}
 \tablefoottext{d}{Watanabe et al. \cite{wa06}.}
 \tablefoottext{e}{Prinja, Barlow \& Howarth \cite{pr90}.}}
   \label{tab:comp}
\end{table*}

In many aspects, 4U\,1907+09 is similar to Vela\,X-1: both systems
host OB supergiants of almost equal mass and radius, the orbital
periods of both systems are almost equal to each other as well,
and both systems are bow shock-producing runaways. Below we
compare the bow shocks associated with these HMXBs.

The geometry of a bow shock generated by a supersonically moving
star is characterized by two main parameters, the Mach number
$M=v_\ast /c_{\rm s}$ (where $v_\ast$ is the velocity of the star
relative to the ambient medium and $c_{\rm s}$ is the sound speed
of the ambient medium) and the stand-off distance $R_{\rm s}$. The
higher $M$ (or the higher $v_\ast$) the smaller the opening angle,
$\alpha$, of the bow shock, which is related to the Mach number
through the relationship: $\alpha =2\arcsin (1/M)$. In practice,
however, this angle is difficult to determine owing to the low
emission measure of the flanks of bow shocks. Still, a comparison
of Fig.\,\ref{fig:bow} and Fig.\ref{fig:Vela} suggests that the
opening angle of the bow shock generated by 4U\,1907+09 is much
smaller than that of the bow shock associated with Vela\,X-1,
meaning that the peculiar velocity of the former system is much
larger than that of the latter one (see also below).

Let us estimate the space velocity of Vela\,X-1 using the proper
motion of this HMXB from the new reduction of the {\it Hipparcos}
data by van Leeuwen (\cite{va07}), $\mu _{\alpha} \cos \delta =
-5.48\pm 0.35 \, \masyr, \mu _{\delta} = 8.79 \pm 0.40 \, \masyr$,
and the generally accepted distance to Vela\,X-1 of 1.9 kpc
(Sadakane et al. \cite{sa85}). Using the Galactic constants and
the solar peculiar motion adopted in Sect.\,\ref{sec:run}, we
found the transverse peculiar velocity of Vela\,X-1:
\begin{equation}
v_l = -35.4\pm 3.4 \, \kms , v_b =23.8\pm 3.3 \, \kms \, ,
\nonumber \label{eqn:velo}
\end{equation}
or $v_{\rm tr} = 42.7\pm 3.4 \, \kms$ (cf. Moffat et al.
\cite{mo98}; Huthoff \& Kaper \cite{hu02}). To this velocity one
should add the peculiar radial velocity $v_{\rm rad} =-27 \,
\kms$, inferred from the heliocentric one of $-4 \, \kms$ (Gies \&
Bolton \cite{gi86}), so that the total (three-dimensional)
velocity of Vela\,X-1 $v_\ast \simeq 50 \, \kms$, and its vector
is inclined to the plane of the sky by an angle of $\simeq
32\degr$.

The second parameter characterizing a bow shock is the stand-off
distance, i.e. the minimum distance from the star at which the ram
pressure of the stellar wind is balanced by the ram pressure of
the ambient interstellar medium (e.g., Baranov, Krasnobaev \&
Kulikovskii \cite{ba71}):
\begin{equation}
R_{\rm s} \, = \, \left({\dot{M} v_{\rm w} \over 4\pi \rho v_\ast
^2}\right)^{1/2} \, , \label{eqn:stand}
\end{equation}
where $\dot{M}$ and $v_{\rm w}$ are the stellar mass-loss rate and
the wind velocity, $\rho =1.4m_{\rm H} n$, $m_{\rm H}$ is the mass
of a hydrogen atom, and $n$ is the number density of the ambient
medium. $R_{\rm s}$ is related to the observed angular separation
between the apex of the bow shock and the star, $\theta$, through
the relationship:
\begin{equation}
\theta = \beta \gamma (i) R_{\rm s} /d \, ,
\end{equation}
where $\beta > 1$ is a factor of order unity taking into account
the finite thickness of the layers occupied by the shocked gas
coming from the ambient medium and the stellar wind [ see, e.g.,
Fig.\,1 in Comer\'{o}n \& Kaper (\cite{co98}) for schematic of the
multi-layer structure of a wind bow shock],
\begin{equation}
\gamma (i) = {1 \over \sin i} \sqrt{3\left(1-{i \over \tan
i}\right)} \label{eqn:gamma}
\end{equation}
is a geometric factor taking into account the inclination of the
bow shock to the plane of the sky (see Wilkin \cite{wi96}), where
$i=\arctan (v_{\rm rad} /v_{\rm tr})$ is the angle between the
vector of the space velocity of the star and the plane of the sky.
Note that for small inclination angles ($i<45\degr$) the factor
$\gamma$ is $\simeq 1$.

Using Eqs\,(\ref{eqn:stand})-(\ref{eqn:gamma}), one can estimate
the number density of the ambient medium:
\begin{eqnarray}
n \simeq 0.5 \, {\rm cm}^{-3} \, (\beta \gamma )^2 \left({\dot{M}
\over 10^{-6} \, M_{\odot} {\rm yr}^{-1}}\right) \left({v_{\rm w}
\over 2000 \,
{\rm km } \, {\rm s}^{-1}}\right)  \ \nonumber \\
\times \left({R_{\rm obs} \over 1 {\rm pc}}\right)^{-2}
\left({v_{\ast} \over 100\, \kms}\right)^{-2} \, , \label{eqn:den}
\end{eqnarray}
where $R_{\rm obs} =\theta d$. Inserting parameters of 4U\,1907+09
and Vela\,X-1 (given in Table\,\ref{tab:comp}) in
Eq.\,(\ref{eqn:den}), one finds the number density of the
interstellar medium surrounding these HMXBs, $\simeq 2$ and $10 \,
{\rm cm}^{-3}$, respectively (we assumed here that $\beta \gamma
\sim 1$). Note that despite the much larger separation of
Vela\,X-1 from the Galactic plane ($\simeq 130$ pc versus $\simeq
30$ pc for 4U\,1907+09) it is located in a denser medium than
4U\,1907+09, implying that Vela\,X-1 met a density enhancement (a
cloudlet) on its way (cf. Gvaramadze \& Bomans \cite{gv08};
Gvaramadze et al. \cite{gva09}). This implication is supported by
the presence of an extended (of angular radius of $\sim 0\fdg2$)
{\it IRAS} source at the position of Vela\,X-1, suggesting that
this source might be a Str\"{o}mgren zone created by the
ultraviolet emission of Vela\,X-1 (Kaper et al. \cite{ka97}).
Assuming that the gas in the H\,{\sc ii} region is fully ionized
and its temperature is $10^4$ K, one can obtain an estimate of the
gas number density (e.g. Lequeux \cite{le05})
\begin{equation}
n \simeq 15 \, {\rm cm}^{-3} \, \left({R_{\rm St} \over 7 \, {\rm
pc}}\right)^{-3/2} \left({S(0) \over 3\times 10^{48} \, {\rm
photons} \, {\rm s}^{-1}}\right)^{1/2} \, , \label{eqn:str}
\end{equation}
where $R_{\rm St}$ is the Str\"{o}mgren radius (derived from the
angular radius of the H\,{\sc ii} region and the distance to
Vela\,X-1 of 1.9 kpc) and $S(0)$ is the total ionizing-photon
luminosity of an B0.5\,Iab star (cf. Martins, Schaerer \& Hillier
\cite{ma05}). The good agreement between the two independent
estimates derived from Eqs\,(\ref{eqn:den}) and (\ref{eqn:str})
supports the interpretation of the extended infrared emission
around Vela\,X-1 and its bow shock as the H\,{\sc ii} region
[Kaper et al. \cite{ka97}; see also Fig.\,5 in Gvaramadze \&
Bomans (\cite{gv08}) for another example of a bow shock within an
H\,{\sc ii} region].

It should be noted that the geometric factor $\gamma (i)$ [given
by Eq.\,(\ref{eqn:gamma})] was derived in the thin-shell
approximation, i.e. under the assumption that the thickness of the
region occupied by the shocked gas is $<< R_{\rm s}$ (Wilkin
\cite{wi96}). This approximation, however, could be valid only for
bow-shocks generated in the very dense medium (e.g. cometary
ultra-compact H\,{\sc ii} regions; Mac Low et al. \cite{ma91}) or
for bow shocks produced by stars with dense, slow winds (e.g. red
supergiants). In both cases, the shocked gas could be dense enough
to rapidly cool and collapse into a thin shell. The thin-shell
approximation hardly can be used to describe the shape of
adiabatic bow shocks, i.e. the bow shocks generated by massive
stars possessing fast, hot winds and/or moving through the space
with high peculiar velocities. For these bow shocks, the thickness
of the layers occupied by the shocked stellar wind (the inner
layer) and the shocked ambient medium (the outer layer)
constitutes a significant fraction of $R_{\rm s}$ (Raga et al.
\cite{ra97}; Comer\'{o}n \& Kaper \cite{co98}).

Because 4U\,1907+09 and Vela\,X-1 both contain supergiant stars
with strong winds (see Table\,\ref{tab:comp}), one can expect that
at least the inner layer of the bow shocks generated by these
HMXBs is adiabatic. To check this, we use the following
back-of-the-envelope estimates. The shocked gas in a bow shock
will collapse into a thin shell if the characteristic time-scale
for radiative cooling of the gas,
\begin{equation}
t_{\rm cool} \simeq {3kT \over n_{\rm ps} \Lambda (T)} \, ,
\label{eqn:cool}
\end{equation}
is much shorter than the characteristic dynamical time-scale of
the bow shock,
\begin{equation}
t_{\rm dyn} \sim {R_{\rm s} \over v_{\ast}} \, , \label{eqn:dyn}
\end{equation}
where $k$ is the Boltzmann constant, $T_{\rm ps}$ and $n_{\rm ps}$
are the initial post-shock temperature and number density, and
\begin{equation}
\label{mcore} \Lambda (T)=\left\{
\begin{array}{l}
0, T<10^4 \, {\rm K} \\
1.0\times 10^{-24} T^{0.55} , 10^4 \, {\rm K} <T<10^5 \, {\rm K}   \\
6.2\times 10^{-19} T^{-0.6} , 10^5 \, {\rm K} <T<4\times 10^7 \, {\rm K}  \\
2.5\times 10^{-27} T^{0.5} , T>4\times 10^7 \, {\rm K}
\end{array}
\right. \label{eqn:lam}
\end{equation}
is the cooling function (Raymond, Cox \& Smith \cite{ra76}; Cowie,
McKee \& Ostriker \cite{co81}). For the (inner) layer of the bow
shock occupied by the shocked stellar wind, $T_{\rm ps} \simeq 1.4
\times 10^7 (v_{\rm w} /1000 \, \kms )^2$ and $n_{\rm ps} =4n_{\rm
w}$, where $n_{\rm w} \simeq \dot{M}/(5.6\pi m_{\rm H} R_{\rm s}
^2 v_{\rm w})$. Using Eqs.\,(\ref{eqn:cool})-(\ref{eqn:lam}) and
parameters of the stellar winds from Table\,\ref{tab:comp}, one
finds that $t_{\rm cool} >> t_{\rm dyn}$ for both bow shocks, i.e.
the inner layer of the bow shocks is adiabatic. Similarly,
adopting for the shocked ambient gas $T_{\rm ps} \simeq 1.4 \times
10^5 (v_{\ast} /100 \, \kms )^2$ and $n_{\rm ps} =4n$, one finds
that $t_{\rm cool} \sim t_{\rm dyn}$ for 4U\,1907+09 and $t_{\rm
cool} << t_{\rm dyn}$ for Vela\,X-1, so that the outer layer of
the bow shock produced by 4U\,1907+09 is adiabatic, while that of
the bow shock around Vela\,X-1 is radiative. From this it follows
that the shape of both bow shocks cannot be described by the
thin-shell solution of Wilkin (\cite{wi96}). Indeed, the Wilkin's
solution predicts that the radial extent of a bow shock in the
direction perpendicular to the direction of motion $R(\pi
/2)=\gamma (\pi /2) R_{\rm s} \simeq 1.7 R_{\rm s}$, while from
Fig.\,\ref{fig:bow} and Fig.\,\ref{fig:Vela} we found $R(\pi
/2)\simeq 1.5 R_{\rm s}$ for 4U\,1907+09 and $R(\pi /2)\simeq 2.4
R_{\rm s}$ for Vela\,X-1.

It is also instructive to compare the geometry of the bow shocks
produced by 4U\,1907+09 and Vela\,X-1 with that of model bow
shocks from Comer\'{o}n \& Kaper (\cite{co98}), namely with their
case D and case E bow shocks (generated by stars with space
velocities of 50 and $150 \, \kms$, respectively, which are almost
identical to those of Vela X-1 and 4U\,1907+09). One can show that
like in the case of 4U\,1907+09 and Vela\,X-1, the inner layer of
both model bow shocks is adiabatic (cf. Comer\'{o}n \& Kaper
\cite{co98}), while the outer one is radiative in the case D bow
shock and adiabatic in the case E bow shock. It is not surprising,
therefore, that the bow shock around Vela\,X-1 shows a close
similarity to the case D bow shock, while the shape of the bow
shock associated with 4U\,1907+09 reminds that of the case E bow
shock [cf. Fig.\,\ref{fig:Vela} and Fig.\,\ref{fig:bow} with the
upper and the lower panels of Fig.\,13 in Comer\'{o}n \& Kaper
(\cite{co98}), respectively]. The more thorough study of the bow
shocks associated with 4U\,1907+09 and Vela\,X-1 requires
numerical simulations [similar to those carried out by Comer\'{o}n
\& Kaper (\cite{co98})], which is beyond the scope of the present
paper.

The inference that the outer layer of the bow shock around
Vela\,X-1 is radiative is consistent with the observational fact
that just this of the two bow shocks manifests itself in
H$\alpha$. On the other hand, the non-detection of the H$\alpha$
emission from the bow shock associated with 4U\,1907+09 could be
used to somewhat constrain the distance to this HMXB. The space
velocity of 4U\,1907+09 approximately scales with the distance as
$\sim d$, so that $n\sim d^{-4}$, and correspondingly, $t_{\rm
cool} \propto d^{7.2}$ (for $10^5 \, {\rm K} <T<4\times 10^7 \,
{\rm K}$) and $t_{\rm dyn} \propto d^0$. The strong dependence of
$t_{\rm cool}$ on $d$ implies that the outer layer of the bow
shock would be radiative, while the bow shock itself would be
visible in H$\alpha$ if the distance to 4U\,1907+09 is shorter
than 4 kpc, say comparable to that of Vela\,X-1 ($\simeq 2$ kpc).
This implication, however, suffers from the large error bar in the
proper motion measurements, which causes the correspondingly large
uncertainties in estimates of the space velocity of the HMXB and
the cooling time-scale of the bow shock. And on the related note,
for distances much larger than 4 kpc, the outer layer of the bow
shock would remain adiabatic, while the space velocity of the
system would become too high to be explained within the
binary-supernova scenario (see Sect.\,\ref{sec:vel}).

Finally, we note that the IRAC 8\,$\mu$m image of the bow shock
produced by Vela\,X-1 (right panel of Fig.\,\ref{fig:Vela}) shows
filamentary (cirrus-like) structures beyond the main body of the
bow shock. We speculate that these filaments are due to
interstellar dust grains aligned with the local interstellar
magnetic field and heated by the radiation of Vela\,X-1.

\section{Summary and conclusion}
\label{sec:sum}

We have searched for bow shocks around 5 HMXBs (GX\,301-2,
4U\,1907+09, Cyg\,X-1, Vela\,X-1 and V615\,Cas) using the {\it
Spitzer Space Telescope} archival data. We found two bow shocks,
one of which (produced by Vela\,X-1) was already known, while the
second one (generated by 4U\,1907+09) was detected for the first
time. The detection of the bow shock around 4U\,1907+09 provides
strong evidence that this HMXB is a runaway system, supporting the
general belief that most of HMXBs possess high space velocities
caused by supernova explosion of one of the components of the
progenitor massive binary. To prove the runaway nature of
4U\,1907+09, we have measured its proper motion using archival
data over a total time baseline of 50 yr. Although suffering from
large error bars, our measurements support the runaway
interpretation of 4U\,1907+09, and suggest that this system was
expelled from the Galactic plane about $10^5$ yr ago. Using the
observed parameters of 4U\,1907+09, we found that the high space
velocity of this HMXB could be well explained within the framework
of the symmetric binary-supernova explosion. We found also that
the possible asymmetry in the supernova explosion does not
contribute significantly to the space velocity of 4U\,1907+09. We
have compared the bow shocks produced by 4U\,1907+09 and Vela\,X-1
and showed that the cooling of the shocked stellar wind is
inefficient in both shocks. We showed also that the layer of the
shocked ambient gas is adiabatic in the former bow shock and is
radiative in the latter one. This allowed us to explain
qualitatively why the shapes of both bow shocks do not follow the
thin-shell solution by Wilkin (\cite{wi96}).

To conclude, we expect that the future proper motion and parallax
measurements for 4U\,1907+09 with the space astrometry mission
{\it Gaia} will considerably improve the estimate of its space
velocity, and thereby will allow us to better understand the
evolution and pre-supernova configuration of its progenitor binary
system. With the new proper motion and distance measurements in
hands we will also be able to refine the parameters of the bow
shock around 4U\,1907+09 and to confront them with numerical
simulations.

\begin{acknowledgements}
We are grateful to the referee for comments allowing us to improve
the presentation and the content of the paper. VVG acknowledges
financial support from the Deutscher Akademischer Austausch
Dienst. This research has made use of the NASA/IPAC Infrared
Science Archive, which is operated by the Jet Propulsion
Laboratory, California Institute of Technology, under contract
with the National Aeronautics and Space Administration, the SIMBAD
database, and the VizieR catalogue access tool, both operated at
CDS, Strasbourg, France.
\end{acknowledgements}

\end{document}